\documentclass[a4paper,twocolumn]{aastex6}
\pdfoutput=1
\usepackage{amstext}
\usepackage{amsmath}
\usepackage{amssymb}
\usepackage{natbib}

\newcommand{\msun}{$M_{\odot}\,$}
\newcommand{\mcool}{$M_{\rm cool}\,$}
\newcommand{\beq}{\begin{equation}}
\newcommand{\eeq}{\end{equation}}
\newcommand{\rp}{$r-$process }

\begin{document}

\shortauthors{Macias \& Ramirez-Ruiz}

\title{A Stringent  Limit on the  Mass Production Rate of \lowercase{$r$}-Process Elements in the Milky Way}

\author{Phillip Macias\altaffilmark{1} and Enrico Ramirez-Ruiz\altaffilmark{1}}

\altaffiltext{1}{Department of Astronomy and Astrophysics, University of California, Santa Cruz, CA 95064}

\begin{abstract} 
We analyze data from several studies of metal-poor stars in the Milky Way, focusing on both strong (Eu) and weak (Sr) $r$-process elements. Because these elements were injected in an explosion, we calculate the mass swept up when the blast wave first becomes radiative, yielding a lower limit for the dilution of such elements and hence a lower limit on the ejecta mass which is incorporated into the next generation of stars.  
Our study demonstrates  that in order to explain the largest enhancements  in [Eu/Fe] observed in stars at low  [Fe/H]  metallicities, individual 
$r$-process production events must synthesize a minimum of $10^{-3.5}$ \msun of $r$-process material. We also show that if the site of Mg production is the same as that of Eu, individual injection events must synthesize up to  $\sim 10^{-3}$ \msun  of $r$-process material. On the other hand, demanding that Sr traces Mg production results in \rp masses per event of $\sim 10^{-5 }$ \msun. This suggests  that the astrophysical sites responsible for the genesis of the strong $r$-process elements need to operate at  a drastically  reduced rate when compared to core collapse supernovae, while the synthesis of weak \rp material is consistent with a supernova production site.
\end{abstract}

 \keywords{early universe --- galaxies: high-redshift --- galaxies: evolution --- stars: abundances}

\section{Introduction}\label{sec:intro}
Although the physical conditions required for \rp nucleosynthesis to occur have been understood since  \citet{bbfh} and \citet{cameron1957}, the astrophysical site(s) in which those conditions are realized remains unclear. Whether enrichment has occured via Type II Supernovae \citep[SNe, e.g.][]{woosley1994}, in which the injection in a galaxy occurs frequently ($\sim 10^{-2}$ yr$^{-1}$) with low ($\sim 10^{-5}$ \msun) masses, or through neutron star mergers'  \citep[NSM, e.g.][]{lattimer1974} sporadic ($\sim 10^{-5}$ yr$^{-1}$) injection of high ($\sim 10^{-2}$ \msun) masses is difficult to discern at high metallicities, as any hysteresis has been eradicated by multiple enrichment events. 

For this reason, metal-poor stars in the galactic halo serve as laboratories for the study of $r$-process element synthesis and  can shed light on the identity of their progenitors \citep{sneden2008}. Abundance comparisons between many metal-poor halo stars suggest that the $r$-process mechanism is rather robust. Put differently, we see the same relative proportions of  $r$-process elements in stars that are many  billions of years different in age,  hinting that this process has operated in a fairly consistent manner over the history of the Galaxy. This result has been used  to constrain the specific physical conditions and nuclear properties required for the $r$-process. 

In the  metallicity range  [Fe/H] of roughly -2 to -3.5, where we are using the standard notation [X/H] = log$_{10}$(X/H) - log$_{10}$(X/H)$_\odot$, $r$-process elements have been found to exhibit  large star-to-star bulk scatter in their concentrations with respect to the lighter elements albeit with a distribution that is characteristic of solar system matter. This hints  at the presence of chemically  inhomogeneous and unmixed  gas  at that epoch \citep{fields2002}. As time evolves,  these localized   inhomogeneities are  smoothed out as  more events occur and  $r$-process products  migrate and mix throughout the Galaxy. Recent cosmological simulations of heavy element production in a Milky Way- (MW-)like galaxy have shown the observed stellar abundances resulting from this process to be consistent with NSMs being the dominant enrichment mechanism \citep{shen2015,vandevoort2015}, but must rely on prescriptions regarding how material is mixed in the young MW and suffer from uncertainties in the delay time for NSMs.

 In this {\it Letter} we use  simple and conservative physical arguments to show that the scatter in both strong (Eu) and weak (Sr) \rp elements at low [Fe/H] metallicities can be used to place stringent lower limits on how much $r$-process material  needs to be  synthesized per injection event in the early Universe. In Section \ref{sec:sn} we combine abundance data from several previous studies of MW stars and focus on Mg production to identify stars which may have formed from gas that has been enriched by a single event. In Section \ref{sec:rp} we derive lower limits on the \rp production required to explain Eu enhancements in these same stars, and also demonstrate the implications of demanding that \rp enhancements trace the Mg source. We discuss our findings and conclude in Section \ref{sec:disc}.

\section{Supernova II as testbeds for Metal enrichment}\label{sec:sn}
While there is no current consensus on the dominant channel of $r$-process production, it is well understood that the so called $\alpha$ (O, Mg, Si, Ca, and Ti) elements are primarily produced in massive stars and returned to the ISM via core-collapse SNe \citep{bbfh,woosley1995}. For this reason, elements such as Mg have been measured in metal-poor MW halo stars to study the efficiency of galactic mixing in the early universe \citep{arnone2005}. Here we focus on Mg production in the MW in order to demonstrate how our physical argument applies to a relatively well understood production source. 

SNe input approximately $10^{51}$ erg of energy into their surroundings, resulting in a blast wave which sweeps up a less $\alpha$-enhanced ISM, thereby mixing and diluting any enhancement supplied by the ejecta.  In order to incorporate these metals into a new generation of stars the SN blast wave must first cool, at the very least. The mass swept up before the blast wave becomes radiative and cools efficiently in a homogeneous medium is given by 
\beq
M_{\rm cool} \approx 10^3 \left(\frac{Z}{Z_\odot}\right)^{-3/7} \left(\frac{n_{\rm ISM}}{10^2 \, {\rm cm^{-3}}}\right)^{-2/7}  \left(\frac{E_{\rm exp}}{10^{51} \rm erg}\right)^{6/7}M_\odot ,
\label{eq:mcool}
\eeq
 where $E_{\rm exp}$ is the explosion energy and $Z$ and $n_{\rm ISM} $ are the metallicity and number density of the surrounding ISM, respectively \citep{cioffi1988,thornton1998,martizzi2015}. For a given ejecta mass, the \emph{maximum} enhancement possible of the surrounding ISM (to be observed in the next generation of stars) occurs when the ejecta has mixed with \mcool, as further mixing (which certainly happens due to the inertia of the expanding material as well as   larger scale mixing  due  to e.g.  turbulence generated by galactic shear) will further dilute the enhancement \citep{greif2009}.
  
 One can then invert this relation to find the \emph{minimum mass of the event} for a given enhancement in the next generation of stars, which is given by
\beq
M_{X} \geq X_\odot \times 10\,\,^{{\rm [X/H]}}\,\,M_{\rm cool},
\label{eq:mx}
\eeq
where $X_\odot$ is the mass fraction of element X within the Sun. This is the  mass required to explain a given stellar enhancement, assuming the ejecta is only mixed within a cooling mass. Since the ejecta certainly mixes with greater than a cooling mass and further dilutes, Equation \eqref{eq:mx} represents a strict lower limit for low metallicity stars which have not been enriched by multiple events. Recent simulation work by \citet{martizzi2015} show final swept masses between 1700 \msun in a homogeneous ISM and 8000 \msun in a turbulent ISM with similar scalings, indicating that our analytical estimate is conservative.

\begin{figure}[b]
\plotone{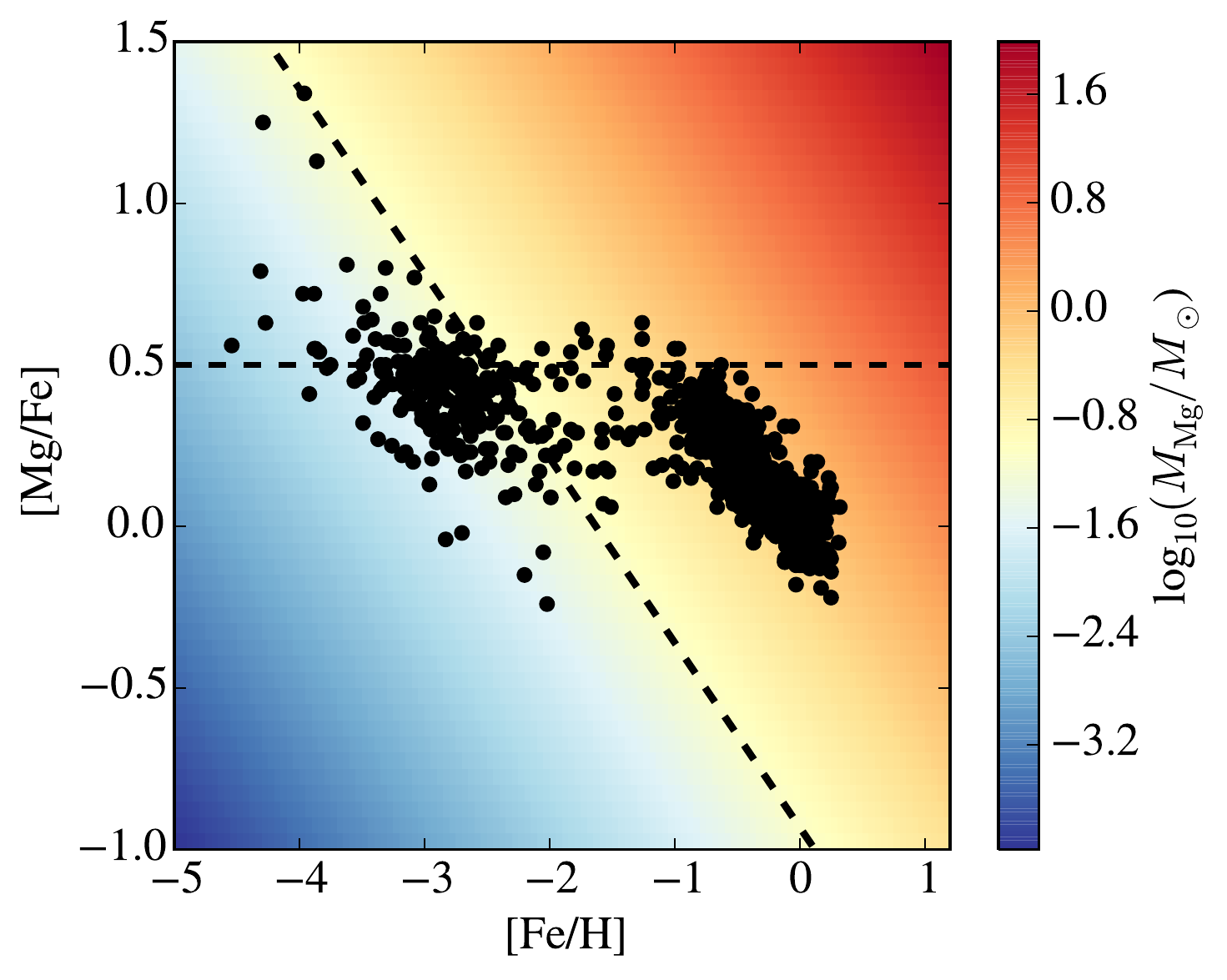}
\caption{Mg abundance as a function of metallicity for our total star sample. In black dots we show abundance data from \citet{roederer2014}, \citet{reddy2006}, \citet{reddy2003}, \citet{cayrel2004}, \citet{venn2004}, \citet{barklem2005} and \citet{fulbright2010}. In color is shown the mass of Mg required to explain the abundances if the ejecta is mixed over one cooling mass (see text). The diagonal line represents a Mg mass of 0.1 \msun and the horizontal line represents the IMF and metallicity-weighted yield for SN ejecta.}
\label{fig:mgfe}
\end{figure}

In Figure \ref{fig:mgfe} we plot [Mg/Fe] as a function of [Fe/H] for a compilation of MW stars and show in color the minimum $M_{\rm Mg}$ required to explain the observations, assuming that the ISM which collapsed to form the stars was enhanced by a single event which input 10$^{51}$ ergs of energy. This simplistic assumption clearly breaks down at high metallicities where the gas has been enhanced by many events over the history of the galaxy, but we note some interesting behavior at low metallicity. First, at [Fe/H] $\lesssim$ -2.5 the stars are all consistent with a minimum Mg mass less than 0.1  \msun, which is shown by the dashed line and is roughly in agreement with the Mg mass expected to be produced in a single SN \citep{nomoto2006,kobayashi2006}. We do not expect a clustering at exactly the dotted line as most SNe likely mix well past their cooling mass, resulting in a vertically downward trajectory on the plot. At higher metallicities, we see a convergence toward [Mg/Fe] = 0.5, which is roughly the IMF-weighted yield of SN ejecta \citep[e.g.,][]{kobayashi2006}.  At this point the gas is well mixed and is clearly incompatible with pollution by a single event as evidenced by the large masses required to explain the enhancement. 

The dearth of stars in the upper-right quadrant can be understood simply. At low metallicities, it is impossible to be polluted by more than the mass of a single event (shown by the horizontal dashed line), while at higher metallicities it is impossible to enrich beyond the yields since mass fraction is an intensive quantity, i.e. the enrichment has saturated to the yields. 

Because we wish to constrain the mass per event of $r$-process material, we do not consider this integrated history and instead focus our attention on metallicities lower than that in which the enriched gas has reached  [Mg/Fe]  abundance ratios close to those given by SN yields.

\begin{figure}
\plotone{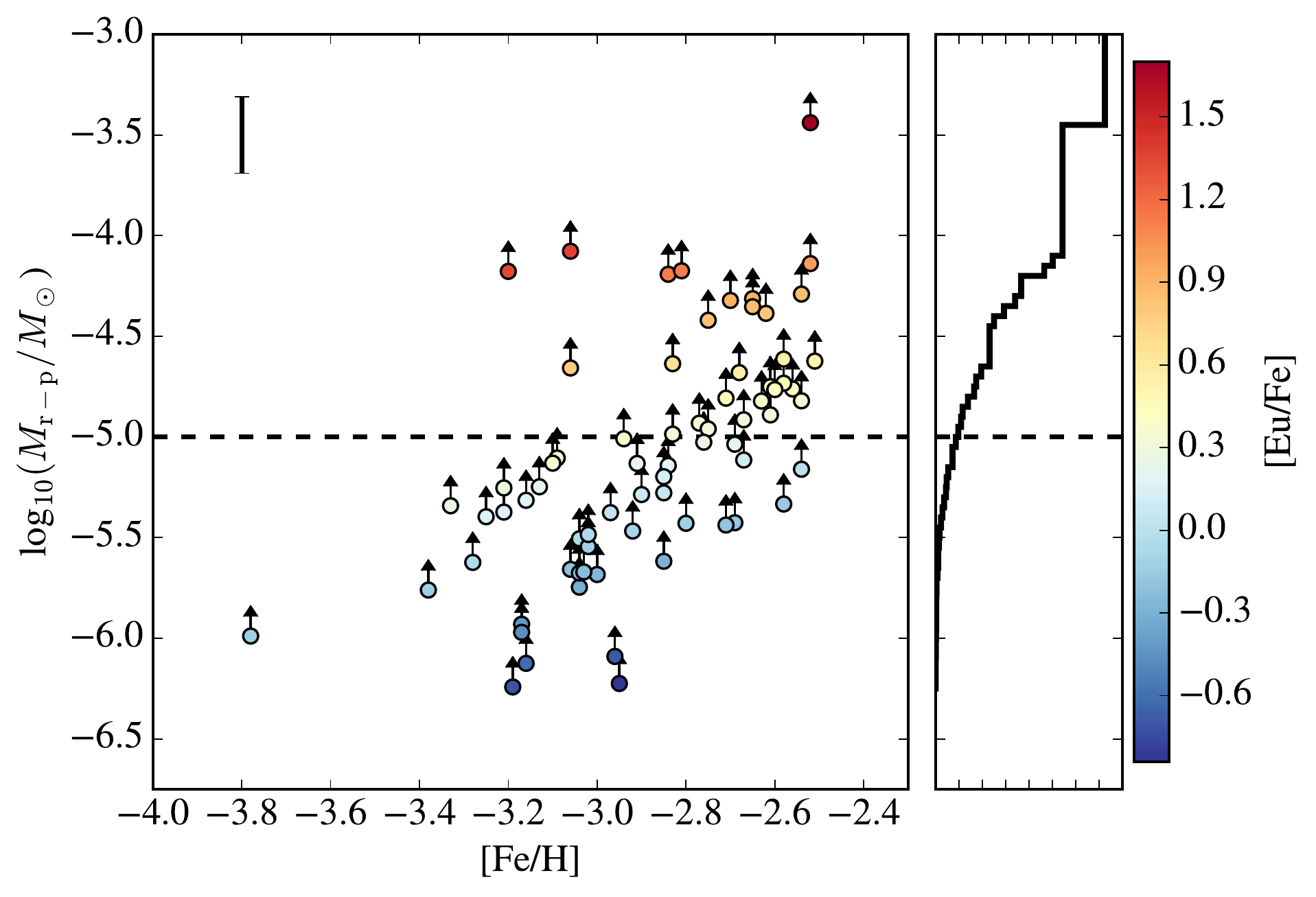}
\caption{Inferred minimum \rp  mass per event as a function of metallicity with a mass-weighted cumulative histogram in projection. Color marks the observed [Eu/Fe] for these stars and the size of the average error bar is shown on the top left. }
\label{fig:eufe}
\end{figure}
\section{Constraints on  \lowercase{$r$}-process production} \label{sec:rp}
\subsection{A Strict Lower Limit}

With the data set now consisting of only these {\it single event candidates}, we can begin to ask more probing questions. First, we can do exactly the same exercise that we did for Mg on a so called strong \rp only element, Eu, for the same set of stars. At these low metallicities, heavier elements such as Eu certainly can not have saturated to the yields, as the overall mass production rate of Mg exceeds that of Eu by several orders of magnitude. In addition, because the \rp pattern has been shown to be robust, we can then scale the Eu mass to a total \rp mass demanded by the Eu abundances. This analysis is not highly dependent on the initial ejecta configuration (e.g. spherical ejecta as opposed to tidal tails), as the initial conditions are quickly forgotten and the blast wave always finds a spherical solution before reaching the cooling radius \citep{montes2016}. 

Figure \ref{fig:eufe} shows the result of this experiment. We find that if the events which caused the Eu enhancement formed the next generation of stars at the cooling mass of the blast wave, these events would correspond to a total \rp mass of up to 10$^{-3.5}$ \msun. We emphasize that each one of these points is a minimum mass per event, and thus represent lower limits. Most of our data are inconsistent with the $10^{-5}$ \msun per event (shown by the diagonal dashed line) necessary to produce the total $r$-process content in the Galaxy given an average SN rate of $10^{-2}$ yr$^{-1}$.  The mass-weighted, cumulative histogram shown in projection on the right can be read as stating that less than 12\% of the \rp mass in the galaxy is consistent with having formed in an event that produced $\le \, 10^{-5}$ \msun.

\subsection{Constraints Based on Mg Mixing}

\begin{figure}
\plotone{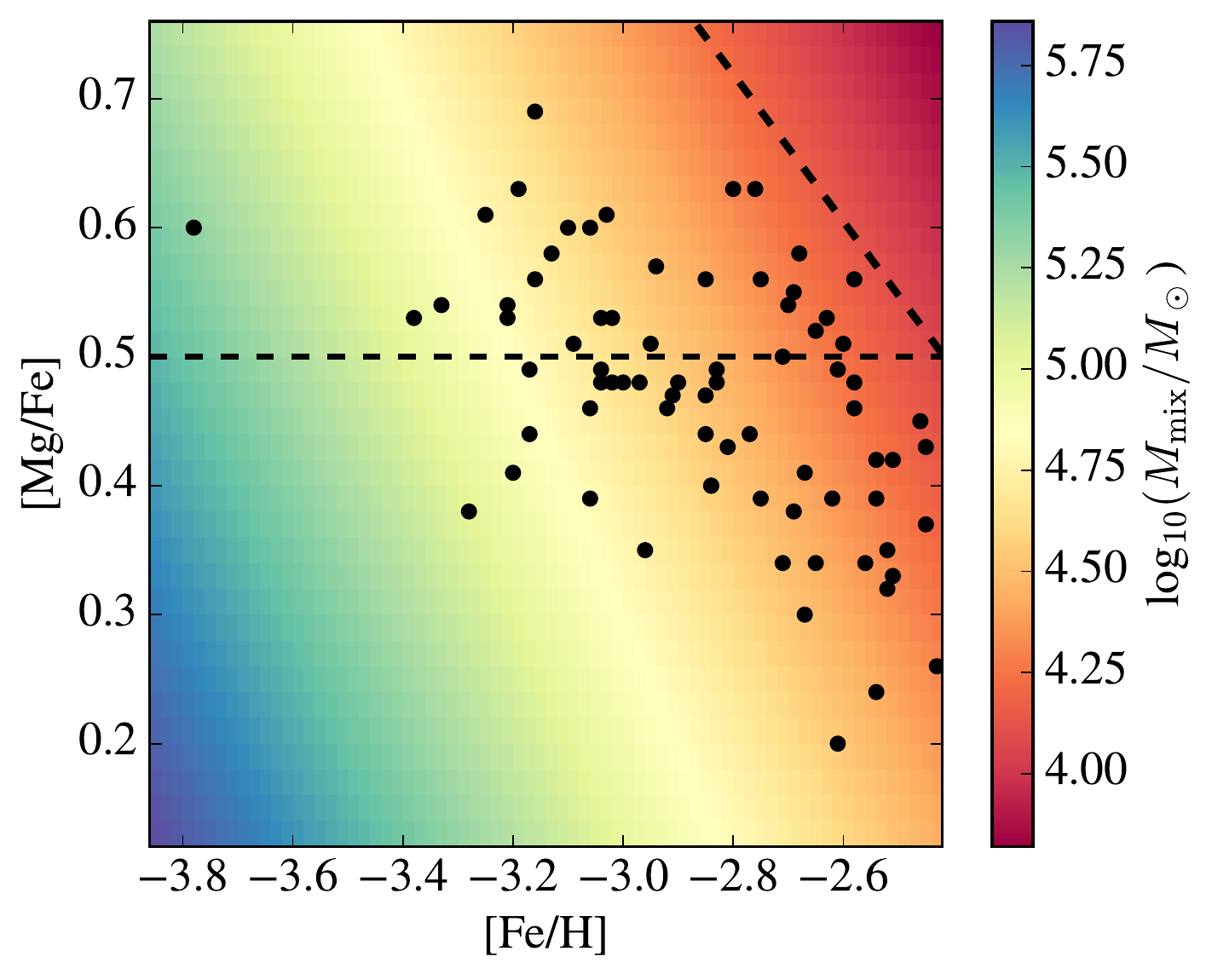}
\caption{ Same as Fig.1, but now showing in color the amount of mass with which our fiducial Mg mass of 0.1 \msun would need to mixed in order to explain the observed stellar abundances for our reduced sample. As metallicity increases and the ISM converges towards the yields, the amount of ISM over which ejecta is enriching decreases.}
\label{fig:mmix}
\end{figure}

Numerical simulations of SN nucleosynthesis have provided us with roughly the total amount of Mg mass ejected in SNe across a wide range of progenitor masses and metallicities \citep[e.g.,][]{kobayashi2006}. Similar to \citet{fields2002}, with these results we can calculate the {\it mixing mass} (denoted here as $M_{\rm mix}$), i.e. the ISM mass over which the Mg must be diluted in order to explain the observed stellar abundances if the subsequent generation of stars were formed by gas which was enriched by a single pollution event,
\beq
M_{\rm mix} = 130 \times 10\,\,^{{\rm- [Mg/H]}}\left(\frac{M_{\rm Mg}}{0.1\,M_{\odot}}\right)\,\,M_{\odot},
\eeq
where we have used $X_{\rm Mg,\odot} = 7.6 \times 10^{-4}$ and a fiducial SN Mg mass of 0.1 \msun. The normalization is not to be taken at face value since the galaxy is well mixed in $\alpha$ elements at [Mg/H] = 0. Figure \ref{fig:mmix} shows our reduced sample, now in color showing $M_{\rm mix}$ for a fiducial SN Mg mass of 0.1 \msun. We can then make the $ansatz$ that the source of the Mg is the same as that of both Eu and Sr. Sr, unlike Eu, is thought to be synthesized in the weak \rp, i.e. in a region of lower neutron to seed ratio. If these elements are coming from the same astrophysical site, this mixing mass should be the same for Eu and Sr as for Mg, allowing us to infer a total r-process mass per event. We convert from an elemental mass to total \rp mass by scaling the relative abundances to match the solar values, i.e.
\beq
M_{\rm r-p} = X_{\rm r-p, \odot} \times 10\,\,^{{\rm [Eu/H]}} M_{\rm mix},
\eeq
for the strong \rp elements, and 
\beq
M^{\rm w}_{\rm r-p} = X^{\rm w}_{\rm r-p, \odot} \times 10\,\,^{{\rm [Sr/H]}} M_{\rm mix},
\eeq
for the weak elements, where we use $X_{\rm r-p, \odot} = 2 \times 10^{-7}$ and  $ X^{\rm w}_{\rm r-p, \odot} = 0.1 \times X_{\rm r-p, \odot}$ is the fraction of total \rp mass in the weak component. 

\begin{figure}
\plotone{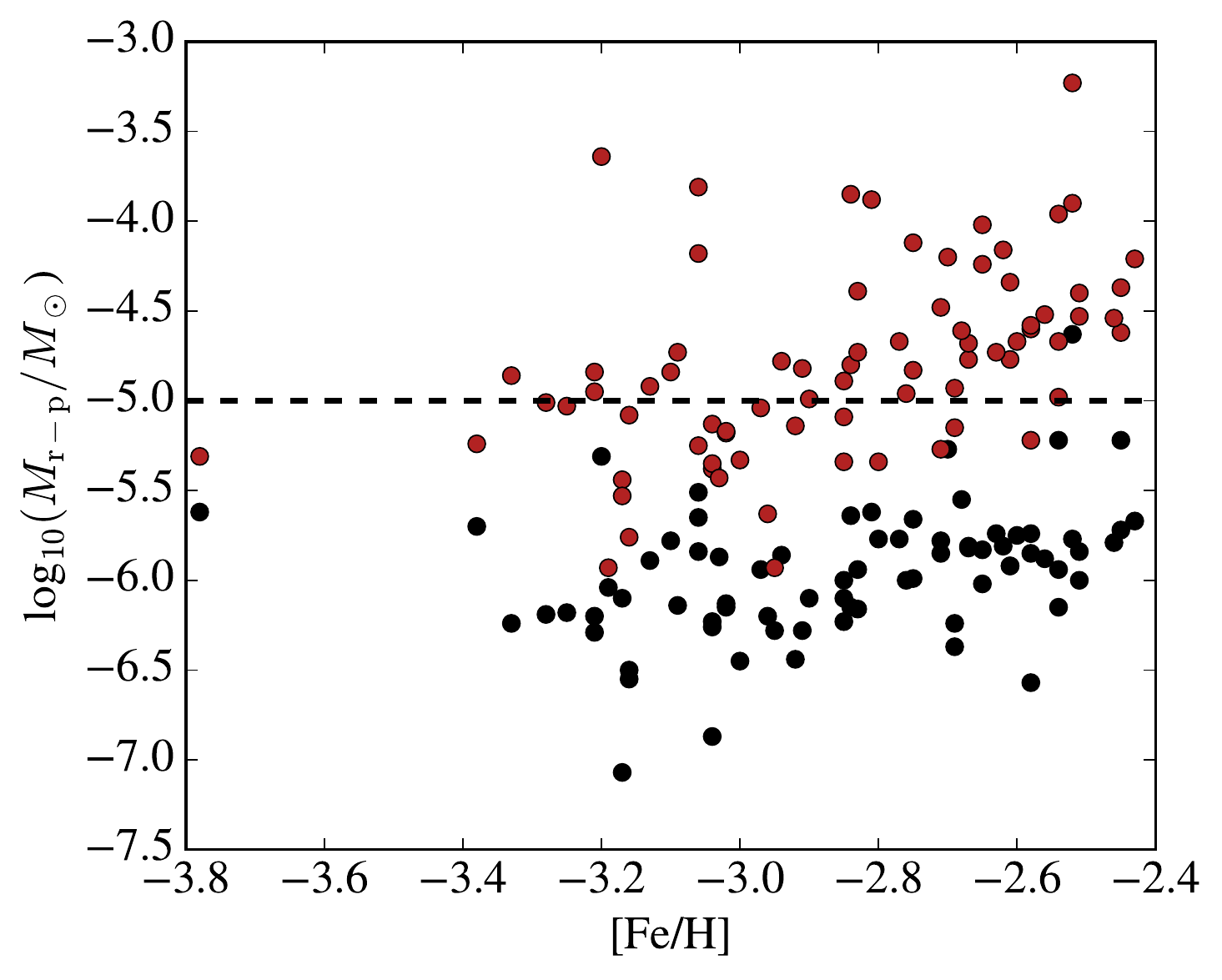}
\caption{Total \rp mass per event required to explain the stellar abundances assuming it mixes over the same mass as the Mg as a function of metallicity. Limits for Eu are shown as  {\it red} symbols while {\it black} symbols show the limits for Sr.}
\label{fig:eu_sr}
\end{figure}

 Mixing is element-agnostic, so we can test the consequences of this hypothesis and answer the question of $r$-process production required by SN in order to explain the observed abundances at low metallicities. Figure \ref{fig:eu_sr} shows the results. For Sr, a weak \rp element, the total \rp mass inferred from the requirement of originating from the same source as Mg is on average $\sim 10^{-6}$ \msun, which is in agreement with the total mass production rate of weak \rp elements in the galaxy given a rate of $10^{-2}$ yr$^{-1}$.

However, demanding that Eu traces the Mg results in total \rp masses well above 10$^{-5}$ \msun. This serves as a proof by contradiction: requiring that the channel providing Mg enrichment in the early universe is the same as that which provides Eu would drastically overproduce the total \rp mass in the galaxy today. 

We note that $M_{\rm mix}$ should be inversely proportional to the rate of injection, i.e. that rarer events will be spread out further in distance as well as time, and will thus dilute further between events, e.g. through turbulent diffusion. In this way, this experiment implies another lower limit on $M_{\rm rp}$. Any event which is rarer implies a mass per event larger than seen in Figure \ref{fig:eu_sr}, and any event with higher rates would overproduce the galactic $r$-process even more drastically.

\section{Discussion}\label{sec:disc}

By looking at metal-poor stars in the MW we are able to place strong constraints on the mass per event and hence rate of the events which have enriched them in \rp elements. As seen in Figure \ref{fig:eufe}, at least 90\% of the total \rp mass in the galaxy must have been synthesized in events that output $> 10^{-5}$ \msun of \rp material, translating to a rate of $< 10^{-2}$ yr$^{-1}$ in order to match the total \rp synthesis rate in the MW of $10^{-7}$ \msun yr$^{-1}$ \citep{cowan2004,sneden2008}. This shows that even under the most conservative assumptions core-collapse SNe are inconsistent with being the dominant progenitor of strong \rp elements in the early universe given their frequency. This analysis is in agreement with several recent arguments, as it is only in the past few years that we have been able to break the degeneracy between rate and mass per event amongst the leading theories by looking further into the history of the galaxy \citep[e.g.,][]{shen2015,ji2016a}.  In addition, we have used a fiducial density of $10^{2}$ cm$^{-3}$ in our calculation of \mcool, whereas NSM are likely to occur in regions of lower density if they receive a kick from the SNe that created the pair \citep[e.g.,][]{belczynski2006,kelley2010}. From Equation \eqref{eq:mcool}, lowering the ambient density by a factor of 100 increases the mass per event by a factor of 4, implying an \rp mass of $\gtrsim 10^{-3}$ \msun per event. 

\citet{beniamini2016} have recently performed a similar analysis using ultra-faint dwarf galaxies (UFDs), assuming a gas mass for the galaxy and calculating the Eu (and hence total \rp) mass required to explain the observed stellar abundances. Their result is in agreement with ours, i.e. they find that the Eu mass per event is inconsistent with enrichment from typical core-collapse SNe given their rate, which naturally extends itself to MW stars assuming the dominant mechanism is the same in both galaxy types. This assumes the ejecta is well mixed throughout the UFD gas, an assumption which we also require at the cooling mass scale, though this is well justified as SN remnants show efficient mixing well before the cooling mass is reached \citep{lopez2011}. Though inhomogeneous mixing may take place at larger scales, this will not re-concentrate a given element. However, our analysis demands an even more conservative lower limit on the \rp mass per event, as our cooling masses are well below the fiducial $10^5$ \msun UFD gas mass. 

Through independent means we are able to look at both the weak and strong \rp elements and calculate the total \rp mass implied by assuming that the source which provided them also generated 0.1 \msun of Mg and scaling the \rp elements to solar abundances. We find that the implied mass per event for strong (Eu) production in most of our stars is $\gtrsim 10^{-5}$ \msun\, and up to $\approx 10^{-3}$ \msun, whereas the majority of weak (Sr) production is consistent with a mass per event of $\lesssim$ 10$^{-5}$ \msun. 

This implies that SNe are consistent with being the dominant source of weak \rp elements in the early universe \citep{surman2014}, and by extension that there may  be two sources of \rp production, consistent with recent findings by \citet{ji2016b}. This argument does not rule out SNe with yields different from typical core-collapse, but any less common supernova must have either a Mg mass greater than 0.1 \msun(to increase the mixing mass) with a rate low enough to not overproduce the total Mg in the galaxy, or a Mg mass much less than 0.1 \msun in order to decouple the Eu production from the Mg. 

\begin{figure}

\plotone{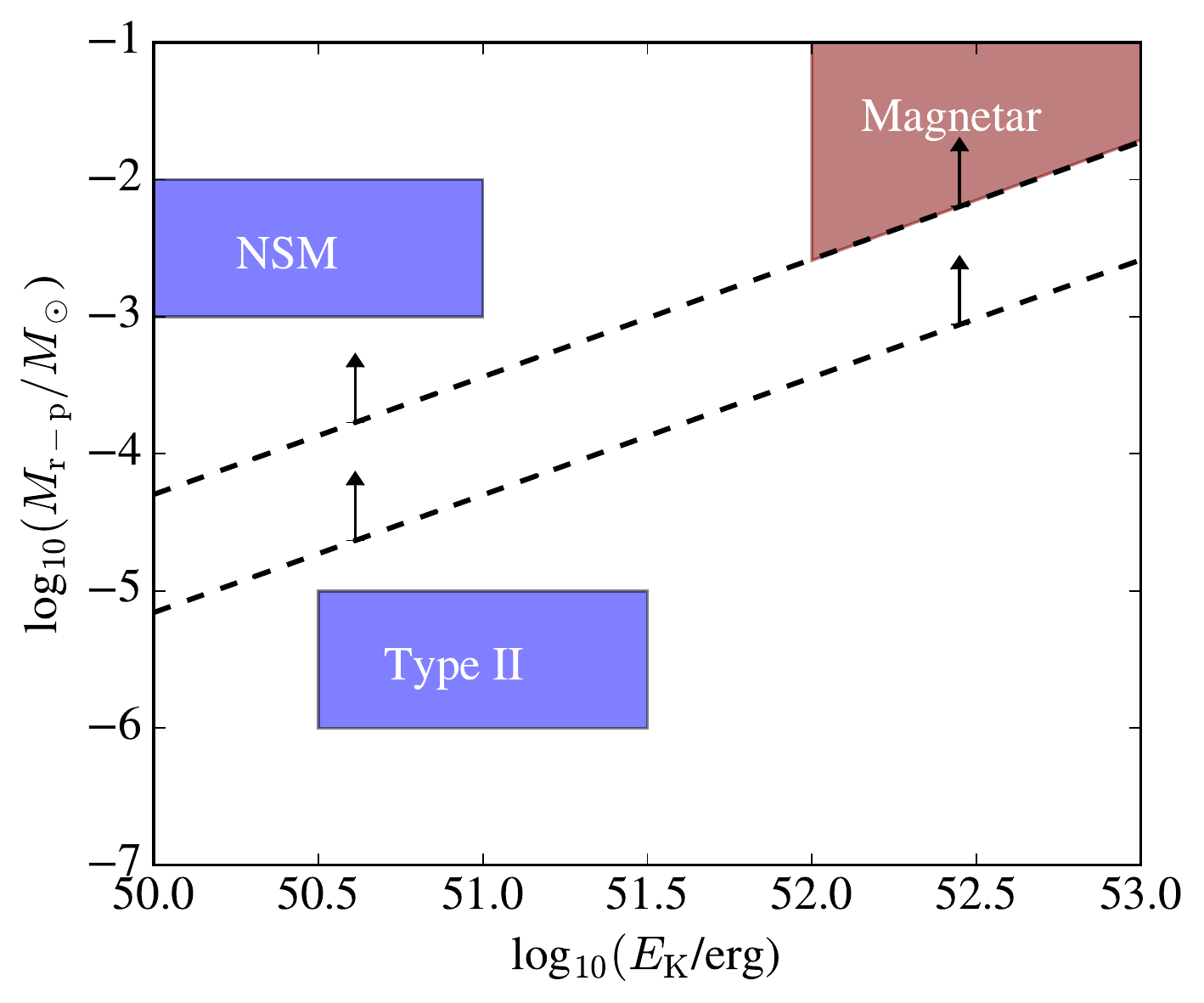}
\caption{Inferred lower limit on \rp ejecta mass based on \mcool from Section \ref{sec:rp}. The dashed lines represent the 100 \% and 50\% values for the mass-weighted cumulative histogram as seen in Figure \ref{fig:eufe}. This argument rules out Type II SNe (purple region denotes the range of current theoretical estimates) as the dominant contributor to the r-process mass budget at low metallicities, and puts constraints on the ejecta mass required in scenarios involving magnetars (maroon region).  }
\label{fig:magnetar}
\end{figure}

The two remaining candidates for the genesis of \rp elements which do not violate these constraints are NSMs \citep[e.g.,][]{lattimer1974,rosswog1999,metzger2010,roberts2011,barnes2013,bauswein2013,grossman2014,ramirez2015}
as well as jet-driven supernovae \citep[e.g.,][]{winteler2012,nishimura2015}, 
both of which are thought to occur less frequently and with larger mass per event, in concordance with this analysis. While we are not able to distinguish between these two, we may be able to place requirements on each scenario by varying the energy of the explosion which provided the enrichment. 

Figure \ref{fig:magnetar} shows how the constraints implied by our cooling mass argument change by varying the energy of the explosion. While we find that SNe are incompatible with any reasonable explosion energy, the energy implied by the spin down of a magnetar in a jet-driven SNe \citep[e.g.,][]{metzger2015} places lower limits on the mass per event of between $> 10^{-3}$ and $> 10^{-2}$ \msun.  Although the data are not yet able to discern between these models, they demand a large mass per event and rate much lower than that of typical type II SNe (based on our cooling mass argument), as well as a Mg mass much greater than 0.1 \msun if the Mg is at all coupled to the Eu (based on our mixing mass argument).

\acknowledgments 
We thank  R. Cooke, D. Kasen, E. Kirby, G. Montes and S. Shen  for insightful discussions and  acknowledge financial support from the Packard Foundation, NSF (AST0847563), UCMEXUS (CN-12-578). PM gratefully acknowledges support from the NSF Graduate Research Fellowship and the Eugene Cota-Robles Graduate Fellowship.

\end{document}